\documentclass[prl,twocolumn,showpacs,preprintnumbers,amsmath,amssymb]{revtex4}

\usepackage{graphicx}
\usepackage{dcolumn}
\usepackage{bm}

\begin{document}


\title{Energy thresholds for discrete breathers}

\author{Michael Kastner}%
\email{kastner@fi.infn.it}%
\affiliation{I.N.F.M., UdR Firenze, Via G.~Sansone 1, 50019 Sesto Fiorentino (FI), Italy}%

\date{\today}

\begin{abstract}
Discrete breathers are time-periodic, spatially localized solutions of the equations of motion for a system of classical degrees of freedom interacting on a lattice. An important issue, not only from a theoretical point of view but also for their experimental detection, are their energy properties. We considerably enlarge the scenario of possible energy properties presented by Flach, Kladko, and MacKay [Phys.\ Rev.\ Lett.\ {\bf 78}, 1207 (1997)]. Breather energies have a positive lower bound if the lattice dimension is greater than or equal to a certain critical value $d_c$. We show that $d_c$ can generically be greater than two for a large class of Hamiltonian systems. Furthermore, examples are provided for systems where discrete breathers exist but do not emerge from the bifurcation of a band edge plane wave. Some of these systems support breathers of arbitrarily low energy in any spatial dimension.
\end{abstract}

\pacs{45.05.+x, 63.20.Pw, 05.45.-a}

\maketitle

The phenomenon of localization is of interest in almost any branch of physics. In addition to the well known Anderson localization due to disorder, the last decade has seen an increasing attention for localization phenomena in translational invariant systems, i.e., in the absence of disorder. Discrete breathers are time-periodic, spatially localized solutions of the equations of motion for a system of classical degrees of freedom interacting on a lattice. A necessary condition for their existence is the nonlinearity of the equations of motion of the system, and the existence of discrete breathers has been proved rigorously for some classes of systems \cite{MacAubBam,AuKoKa,JaNo}. In contrast to their analogues in continuous systems, the existence of discrete breathers is a generic phenomenon, which accounts for considerable interest in these objects from a physical point of view. In fact, recent experiments could demonstrate the existence of discrete breathers in various systems such as coupled optical waveguides \cite{Eisenberg_ea}, low-dimensional crystals \cite{Swanson_ea}, antiferromagnetic materials \cite{SchwarzEnSie}, Josephson junction arrays (\cite{BiUs} and references therein), and molecular crystals \cite{EdHamm}.

Properties of discrete breathers, as well as of some generalizations of discrete breathers, have been studied in detail in a large variety of different models. However, apart from existence proofs and studies of the spatial localization of discrete breathers (which is typically exponential), only a few general results exist. Among these it is worth mentioning the remarkable result by Flach, Kladko, and MacKay \cite{FlaKlaMac} on energy thresholds for discrete breathers in \mbox{one-,} two-, and three-dimensional lattices (and subsequent generalizations to systems with long range interactions \cite{Flach2} and to partially isochronous potentials \cite{DoFla}). These results on energy thresholds have practical relevance, as they can assist in choosing a proper energy range for the detection of discrete breathers in real experiments or computer experiments. The main achievement of these papers on energy thresholds can be summarized as follows: for a certain class of Hamiltonian systems of infinite size, a critical spatial dimension $d_c$ exists, such that
\begin{itemize}
\item for a system whose spatial dimension $d$ is smaller than $d_c$, no positive lower bound on the energy of discrete breathers exist, i.e., discrete breathers of arbitrarily low energy can be found,
\item for a system where $d\geq d_c$, there exists a positive lower bound on the energy of discrete breathers.
\end{itemize}
The critical dimension $d_c$ depends on the type of nonlinearity present in the system and is, according to \cite{FlaKlaMac}, typically two, but never greater than two. This result was obtained by tacitly assuming that the Hamiltonian admits a Taylor series expansion around its equilibrium point. Although this is a reasonable assumption for many systems or models in physics, there exist some remarkable exceptions (see e.g.\ \cite{BaRaChri} and \cite{Nesterenko}). In the present Letter, some assumptions made in \cite{FlaKlaMac} are critically revisited and a larger class of Hamiltonian systems, not necessarily expandable around the equilibrium point, is considered. We find a much richer scenario for the existence of energy thresholds for discrete breathers:
\begin{enumerate}
\item The critical dimension $d_c$ can be greater than two for large classes of Hamiltonian systems. We report on systems which do not show an energy threshold for discrete breathers in spatial dimensions two and three.
\item There exist discrete breathers which do not emerge from a tangent bifurcation of the band edge plane wave. For these discrete breathers an energy threshold exists which, however, cannot be obtained by the analysis used in \cite{FlaKlaMac} and \cite{Flach2,DoFla,Flach1}.
\item In the absence of a linear spectrum (``sonic vacuum''), discrete breathers are superexponentially localized and $d_c=\infty$.
\end{enumerate}

We consider a $d$-dimensional hypercubic lattice with $N$ sites. Each site is labeled by a $d$-dimensional vector $i\in\mathbb{Z}^d$, and a state $x_i\in\mathbb{R}^f$ is assigned to each lattice site, where $f$ is the number of components and is to be finite. The Hamiltonian of the system is of the general form
\begin{equation}
\mathcal{H}(\{x_i\})=\sum_{i\in\mathbb{Z}^d} \left[\mathcal{H}_{\mbox{\small loc}}(x_i) + \mathcal{H}_{\mbox{\small int}}(x_i,\{x_{i+j}\})\right],
\end{equation}
where $\mathcal{H}_{\mbox{\small int}}$ describes the interaction of a state at site $i$ with the states $\{x_{i+j}\}$ in a (finite) neighborhood. $\mathcal{H}$ is assumed to have an equilibrium point at $x_i=0$, with $\mathcal{H}(\{0\})=0$. In order to exemplify, in what follows we shall make reference to the form
\begin{equation}
\mathcal{H}(\{p_i,q_i\})=\sum_{i\in\mathbb{Z}^d} \left[\frac{p_i^2}{2} + V(q_i) + \sum_{j\in N_i}W(q_i,q_j)\right],
\end{equation}
where $V$ is a local potential, $W$ an interaction potential, and $N_i$ denotes the set of nearest neighboring sites of $i$. $V$ and $W$ are assumed to be two times continuously differentiable ($C^2$) and are allowed to have an isolated non-analyticity at zero, being of the shape
\begin{equation}
V(q)=V_2 q^2 + \sum_{n=3}^\infty V_n^\pm q^n,
\end{equation}
where the $V_n$ are constants which may be different for $q\geq0$ ($V_n^+$) and $q<0$ ($V_n^-$). (Analogous for $W$.)

For generic Hamiltonian systems, periodic orbits---and hence also discrete breathers---occur in one-parameter families, and some typical choices of quantities to index such a family are the energy $E$, the amplitude measured at the site with maximum amplitude, or the breather frequency $\omega_b$. It may or may not be the case that, in a certain system, discrete breathers with arbitrarily small amplitude can be found. (This is in contrast to \cite{FlaKlaMac}, where it is argued that the breather amplitude can always be lowered to arbitrarily small values.)

{\em Systems with small amplitude breathers and phonons:} First, it is important to recall that, for discrete breathers to exist as generic solutions, the discrete breather frequency $\omega_b$ has to be nonresonant with the phonon spectrum $\omega_q$, i.e., $\omega_q/\omega_b\notin\mathbb{N}$ \cite{Flach3}. Therefore, in the limit of small amplitudes, the discrete breather frequency $\omega_b$ has to approach an edge of the phonon spectrum from outside the phonon spectrum \cite{FlaKlaMac}. With amplitudes tending to zero, it is observed that localized solutions (discrete breathers) and extended ones (phonons) become more and more similar, and one might suspect that discrete breathers appear through a bifurcation from a phonon mode. For some one-dimensional systems this has been proved \cite{JaNo}, in other cases numerics supports this conjecture. Calculating the critical energy $E_c$ at which this bifurcation occurs, the minimal energy of discrete breathers in a system can be obtained. Flach \cite{Flach1} computed the critical energy for systems where the potentials $V$ and $W$ are infinitely many times differentiable ($C^\infty$) around their minima. Extending this type of analysis to systems where the Hamiltonian is $C^2$ in the vicinity of its equilibrium point, one obtains
\begin{equation}\label{E_c}
E_c \propto N^{1-4/d}
\end{equation}
relating the bifurcation energy $E_c$ to the number of lattice sites $N$. Note that, as $C^\infty$ is a subspace of $C^2$, also the result by Flach, Kladko, and MacKay ($E_c \propto N^{1-2/d}$) can occur, however, as a non-generic case (not robust under $C^2$-perturbations of the Hamiltonian). Additionally to the above mentioned physical applications of nonanalytic potentials, the fact that the generic behavior is somewhat ruled by these potentials renders our result interesting from a theoretical point of view.

Considering the proportionality (\ref{E_c}) in the limit of an infinite number of lattice sites,
\begin{equation}
\lim_{N\to\infty}E_c \left\{
\begin{array}{r@{0\quad\mbox{if}\quad d}c@{4,}}
= & <\\
> & \geq
\end{array}
\right.
\end{equation}
different energy properties are obtained, depending on the spatial dimension $d$ of the system. For $d<4$, discrete breathers of arbitrarily small energies can be found. For $d\geq4$, however, breather energies do not approach zero even if the oscillation amplitude goes to zero, and below a certain positive energy threshold no breathers exist. The value of the threshold energy can be obtained from perturbation theory following the ideas of \cite{Flach1}. So, extending the result of Flach, Kladko, and MacKay \cite{FlaKlaMac}, for this larger class of Hamiltonian systems we obtain a critical dimension $d_c=4$, instead of $d_c=2$, below which generically there is no positive lower bound on the discrete breather energy. 

\begin{figure}[tb]
\includegraphics[width=4.9cm,height=8.8cm,angle=270]{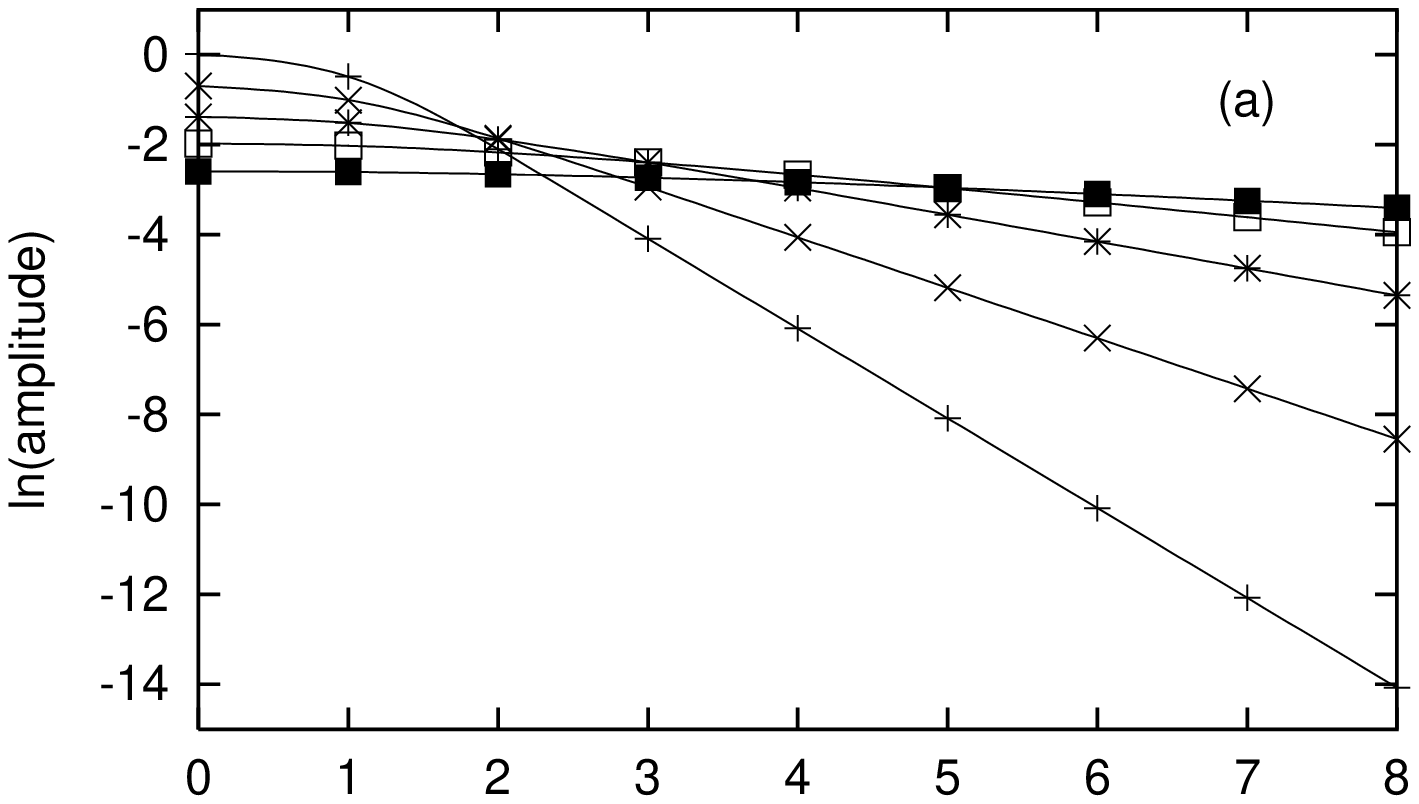}
\includegraphics[width=5.1cm,height=8.8cm,angle=270]{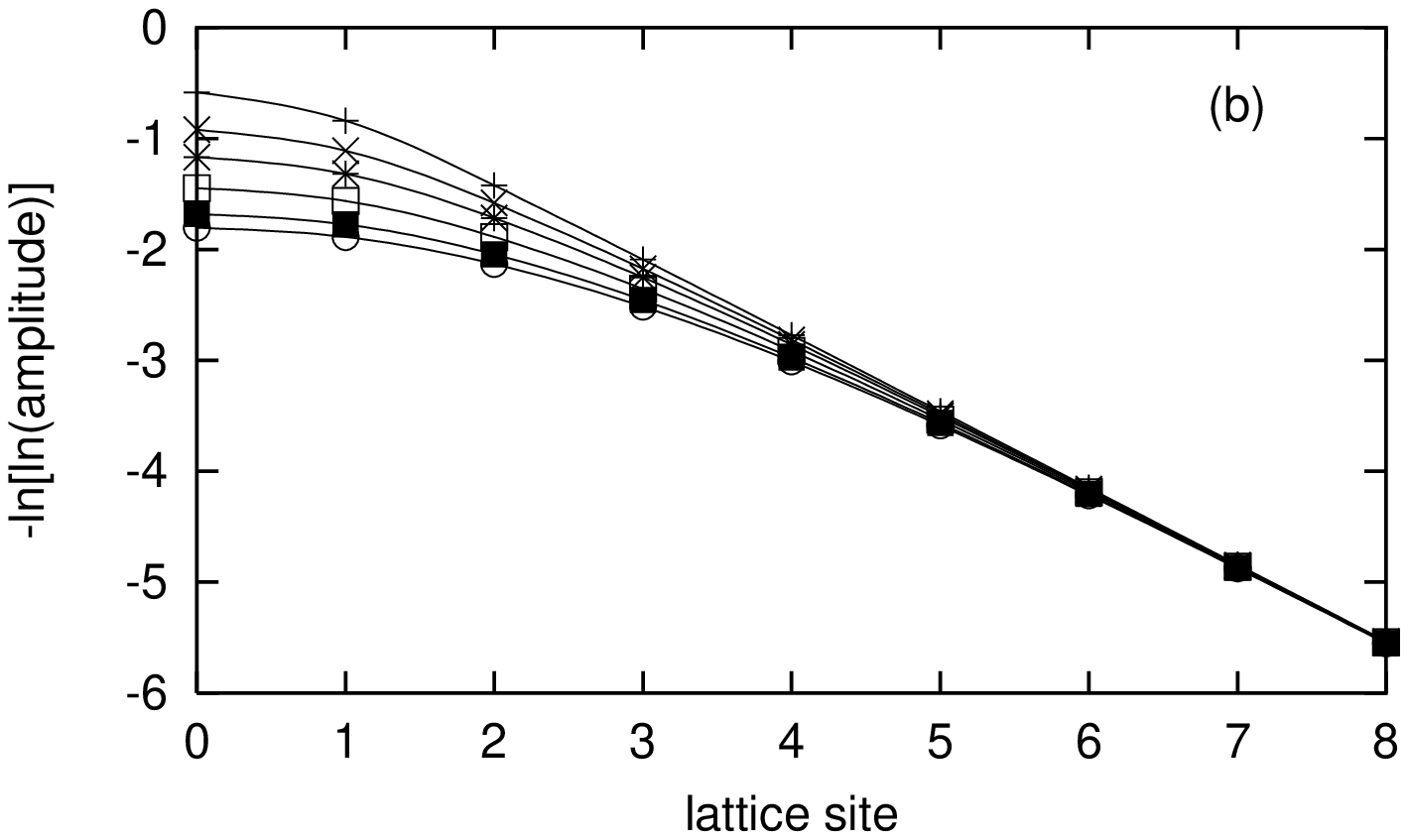}
\caption{\label{DBshape}Shape profiles (absolute value of the amplitude) of discrete breathers, centered at site 0, in Fermi-Pasta-Ulam chains. (a) for a system with phonon spectrum the localization is exponential and decreases when lowering the maximum breather amplitude, while (b), for a system without phonon spectrum, the localization strength is superexponential and remains unchanged. Note the different scales of the ordinates ($\ln(\cdot)$ and $\ln[\ln(\cdot)]$). Lines are merely drawn to guide the eye.}
\end{figure}

The mechanism which leads to the existence or non-existence of an energy threshold can be understood somewhat intuitively in the following way: when lowering the amplitude of a discrete breather towards zero, two different mechanisms, competing with respect to their effect on the breather energy, take place: the localization strength of the discrete breather tends to become weaker, leading to an increase in energy, while the mere lowering in amplitude causes a decrease in energy [see Fig.\ \ref{DBshape}(a) for an example]. Depending on the respective strengths of these effects (which in turn depend on the spatial dimensionality of the system), an energy threshold may exist or not.

\begin{figure}[bt]
\includegraphics[width=4.9cm,height=8.8cm,angle=270]{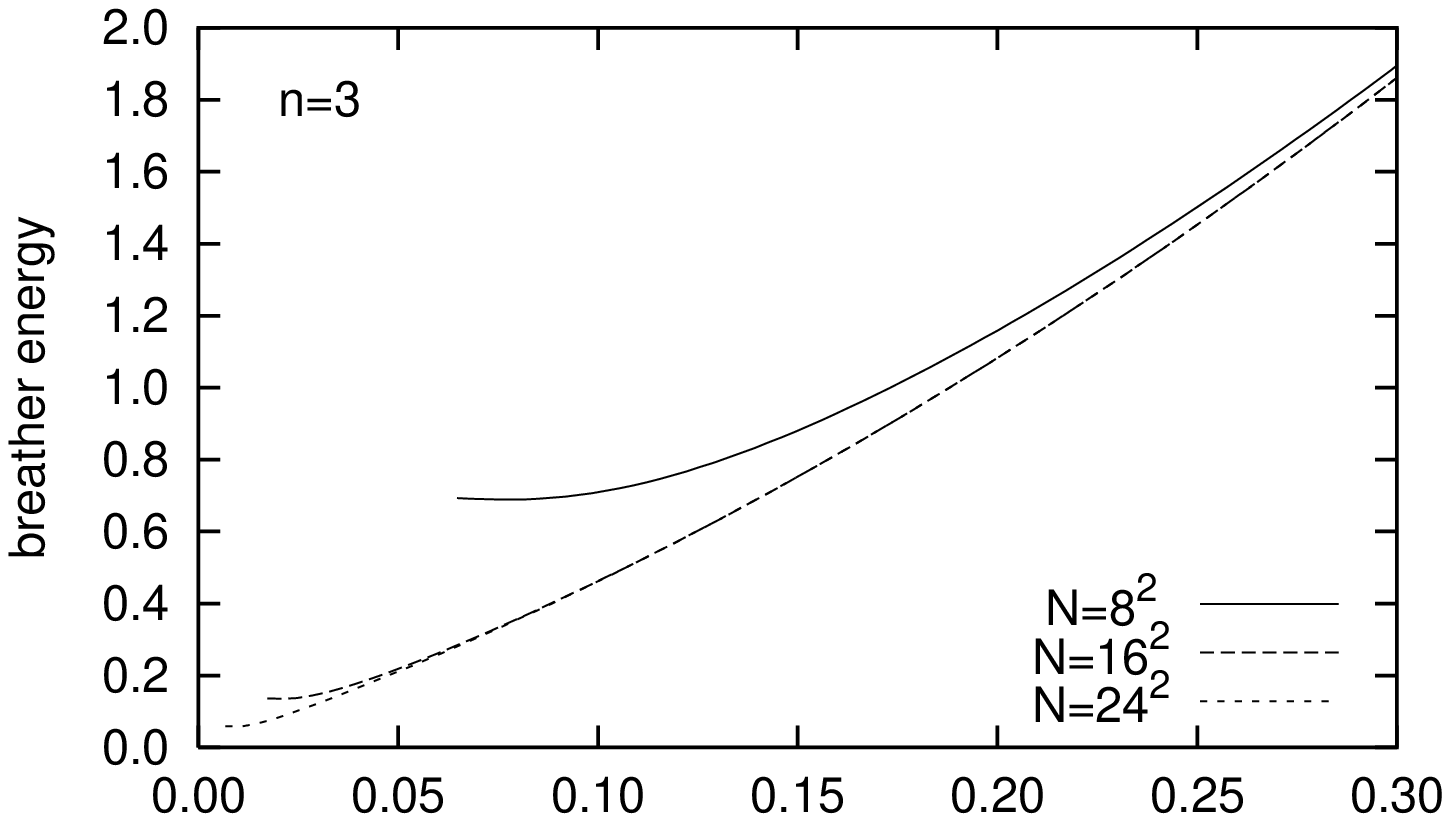}
\includegraphics[width=5.1cm,height=8.8cm,angle=270]{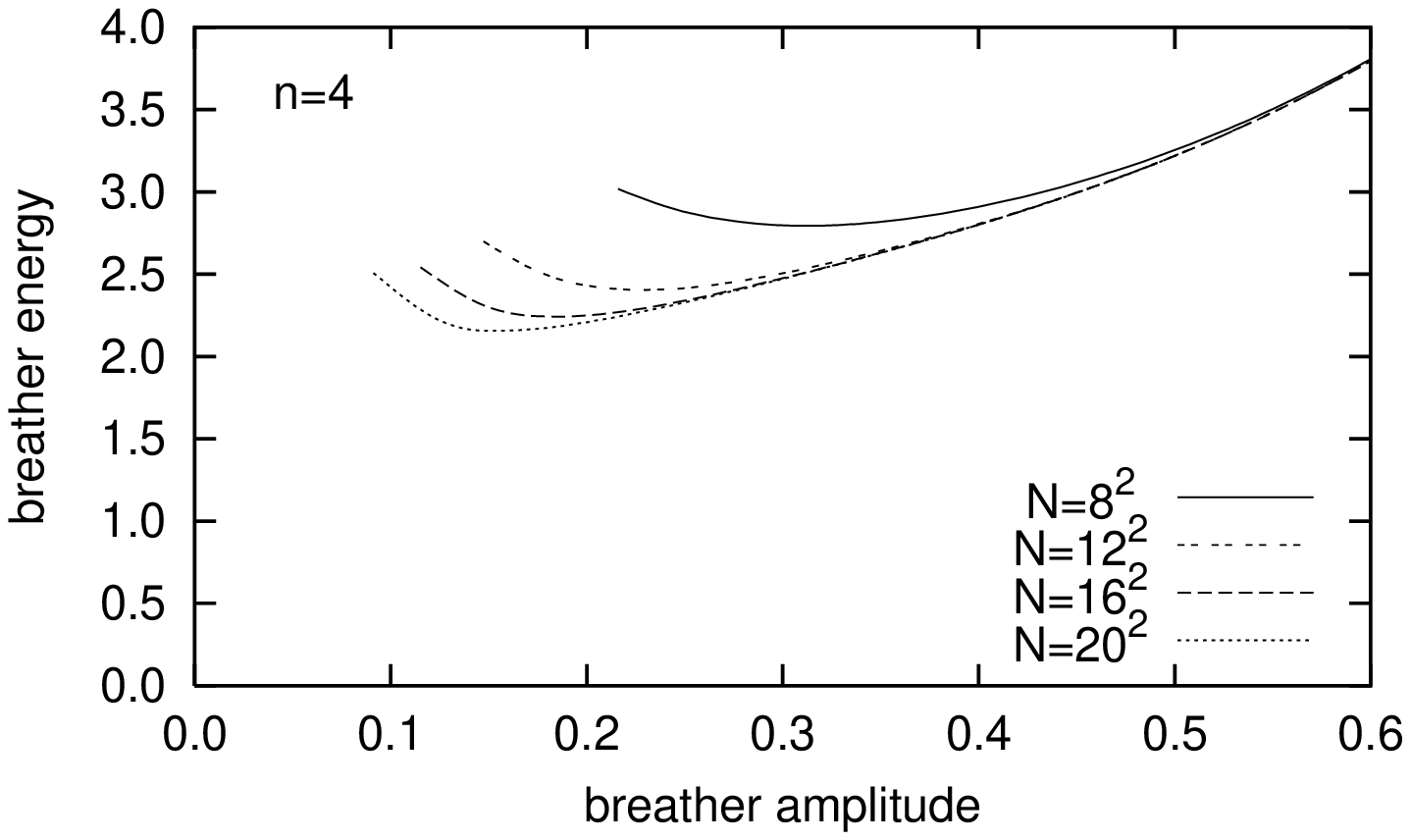}
\caption{\label{en_vs_amp}Breather energy versus amplitude for two-dimensional Fermi-Pasta-Ulam systems with periodic boundary conditions, governed by Eq.\ (\ref{eq_motion}), for the cases $n=3$ and $n=4$. A lower bound on the breather energy is observed. For $n=3$, with increasing system size $N$ this bound converges towards zero, whereas for $n=4$ it converges towards a non-zero value.}
\end{figure}

To confirm the above result, numerical data are presented for systems in two spatial dimensions, the first one showing an energy threshold, the second one supporting breathers of arbitrarily low energy. We consider a Fermi-Pasta-Ulam system in two spatial dimensions with interaction potentials of the form $\frac{1}{2}x^2+\frac{1}{n}|x|^n$ and $n=3,4$. The equations of motion of this Hamiltonian system read
\begin{equation}\label{eq_motion}
\ddot{x}_i = \sum_{j\in N_i}(x_j-x_i)\left[1+|x_j-x_i|^{n-2}\right].
\end{equation}
For $n=4$, the interaction potential, and hence the Hamiltonian $\mathcal{H}$, is $C^\infty$ and we expect the critical dimension to be $2$, whereas, for $n=3$, we have $\mathcal{H}\in C^2\setminus C^3$, and a critical dimension $d_c=4$ is expected. By numerical continuation of periodic orbits from an anti-continuum limit \cite{MaAub}, discrete breathers on finite lattices can be computed numerically up to machine precision. Doing so for a set of frequencies approaching the phonon band edge, the dependence of the breather energy on its amplitude, measured at the site of maximum amplitude, is obtained. The result is plotted in Fig.\ \ref{en_vs_amp}.
For any finite number of lattice sites, a lower bound on the breather energy is observed. For $n=3$, with increasing system size $N$ this bound converges towards zero, whereas for $n=4$ it converges towards a non-zero value, thus confirming the theoretical predictions.

{\em Systems without a phonon spectrum:} A nongeneric, but still remarkable, exception from the above reported energy properties of discrete breathers is observed in systems where no quadratic term is present in the interaction potential. Granular materials provide systems where such interactions are realized and experimentally accessible (see \cite{Nesterenko} for a review). In such systems, discrete breathers have been shown to be superexponentially localized \cite{Flach3,DeyElFlaTsi,Yuan}, where an upper bound on the breather amplitude is given by the inequality $|x_i|\leq a\exp(-|i|/b)c^{(d^i/3)}$ \cite{Yuan}, for some $a,b,c,d\in\mathbb{R}$, in one-dimensional systems. (The breathers are supposed to be centered at site $i=0$.) However, the arguments by which these results for the spatial decay are obtained should be equally applicable in higher spatial dimension, and numerical computations confirm this reasoning. For these sort of systems, the above analysis is not applicable, as discrete breathers do not emerge from the bifurcation of a band edge plane wave. Due to this, and in contrast to the results for systems with phonons reported above, no broadening of the discrete breather takes place when lowering its amplitude [Fig.\ \ref{DBshape}(b)]. It is this constant localization strength---not the superexponential localization!--- which allows for the existence of breathers of arbitrarily low energy {\em regardless of the spatial dimension of the system.}

This is confirmed numerically for a Hamiltonian system with a harmonic on-site potential and an interaction potential of the form $|x|^3$. The equations of motion of this system read
\begin{equation}\label{eq_motion2}
\ddot{x}_i = -x_i+\sum_{j\in N_i}(x_j-x_i)|x_j-x_i|.
\end{equation}
\begin{figure}[bt]
\includegraphics[width=5.1cm,height=8.8cm,angle=270]{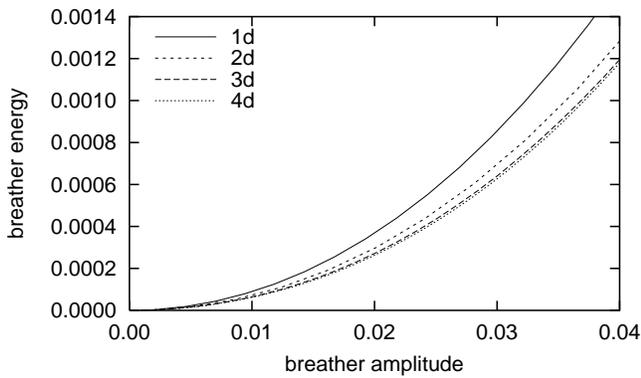}
\caption{\label{en_vs_amp03}Energy versus amplitude for discrete breath\-ers in a system without phonon spectrum [Eq.\ (\ref{eq_motion2})]. Independently of the spatial dimension $d=1,2,3,4$ (from top to bottom), arbitrarily low breather energies are observed. The data were obtained for systems consisting of $N=8^d$ lattice sites, but, due to the strong (superexponential) localization, are indistinguishable on this scale from those of larger systems.}
\end{figure}
\begin{figure}[b]
\includegraphics[width=5.1cm,height=8.8cm,angle=270]{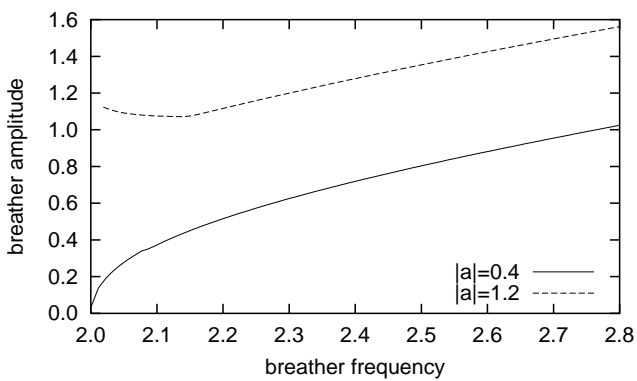}
\caption{\label{amp_vs_freq}Breather amplitude versus frequency for system (\ref{eq_motion3}) with $N=99$ lattice sites, free boundary conditions, and frequencies close to the phonon band edge (which is at frequency $2.0$). With parameter value $|a|=0.4<\frac{1}{2}\sqrt{3}$, breathers of arbitrarily low amplitude are observed, whereas for $\frac{1}{2}\sqrt{3}<|a|=1.2<\sqrt{3}$ the breather amplitude, and hence the breather energy, does not go to zero when approaching the phonon band edge.}
\end{figure}
In Fig.\ \ref{en_vs_amp03}, for systems of spatial dimension $d=1,2,3,4$, the breather energy is plotted versus its amplitude. Independently of $d$, arbitrarily low breather energies are observed already in small finite systems.

{\em Systems without small amplitude breathers:} So far we have considered systems in which the amplitude of a family of discrete breathers can be lowered to arbitrarily small values, and formerly there has been uttered the belief that this were always the case \cite{FlaKlaMac}. From a recent existence proof for discrete breathers by Aubry, Kopidakis, and Kadelburg \cite{AuKoKa} it follows, however, that this is not true in general. In a Fermi-Pasta-Ulam chain with equations of motion
\begin{equation}\label{eq_motion3}
\ddot{x}_i = \sum_{j=i\pm1}[(x_j-x_i)+a(x_j-x_i)^2+(x_j-x_i)^3]
\end{equation}
and parameter values $\frac{1}{2}\sqrt{3}<|a|<\sqrt{3}$, discrete breathers do exist, but their maximum amplitudes lie always above a certain positive lower bound, even when approaching the phonon band edge (Fig.\ \ref{amp_vs_freq}). This automatically implies the existence of a threshold in energy.

\begin{acknowledgments}
Helpful comments from and discussions with Dario Bambusi, J\'er\^ome Dorignac, Sergej Flach, and Roberto Livi are gratefully acknowledged. 
The work was supported by EU contract HPRN-CT-1999-00163 (LOCNET network).
\end{acknowledgments}


\begin{thebibliography}{18}
\bibitem{MacAubBam} R.\ S.\ MacKay and S.\ Aubry, 
Nonlinearity {\bf 7}, 1623 
 (1994); 
D.\ Bambusi, 
{\em ibid.}\/ {\bf 9}, 433 
 (1996).
\bibitem{AuKoKa} S.\ Aubry, G.\ Kopidakis, and V.\ Kadelburg, 
Discrete and Continuous Dynamical Systems: Series B {\bf 1}, 271 
 (2001).
\bibitem{JaNo} G.\ James, 
C.\ R.\ Acad.\ Sci.\ Paris, S\'erie I, {\bf 332}, 581 
 (2001); 
J.\ Nonlinear Sci.\ {\bf 13}, 27 
 (2003); 
G.\ James and P.\ Noble, 
in Proceedings of the Conference on Localization and Energy Transfer in Nonlinear Systems, San Lorenzo de El Escorial, 2002 (unpublished). 
\bibitem{Eisenberg_ea} H.\ S.\ Eisenberg, Y.\ Silberberg, R.\ Morandotti, A.\ R.\ Boyd, and J.\ S.\ Aitchison, 
Phys.\ Rev.\ Lett.\ {\bf 81}, 3383 
 (1998).
\bibitem{Swanson_ea} B.\ I.\ Swanson, J.\ A.\ Brozik, S.\ P.\ Love, G.\ F.\ Strouse, A.\ P.\ Shreve, A.\ R.\ Bishop, W.-Z.\ Wang, and M.\ I.\ Salkola, 
Phys.\ Rev.\ Lett.\ {\bf 82}, 3288 
 (1999).
\bibitem{SchwarzEnSie} U.\ T.\ Schwarz, L.\ Q.\ English, and A.\ J.\ Sievers, 
Phys.\ Rev.\ Lett.\ {\bf 83}, 223 
 (1999).
\bibitem{BiUs} P.\ Binder and A.\ V.\ Ustinov, 
Phys.\ Rev.\ E {\bf 66}, 016603 
 (2002).
\bibitem{EdHamm} J.\ Edler and P.\ Hamm, 
J.\ Chem.\ Phys.\ {\bf 117}, 2415 
 (2002).
\bibitem{FlaKlaMac} S.\ Flach, K.\ Kladko, and R.\ S.\ MacKay, 
Phys.\ Rev.\ Lett.\ {\bf 78}, 1207 
 (1997).
\bibitem{Flach2} S.\ Flach, 
Physica D {\bf 113}, 184 
 (1998).
\bibitem{DoFla} J.\ Dorignac and S.\ Flach, unpublished.
\bibitem{BaRaChri} O.\ Bang, J.\ J.\ Rasmussen, and P.\ L.\ Christiansen, 
Nonlinearity {\bf 7}, 205 
 (1994).
\bibitem{Nesterenko} V.\ F.\ Nesterenko, {\em Dynamics of Heterogeneous Materials} (Springer, Berlin, 2001).
\bibitem{Flach1} S.\ Flach, 
Physica D {\bf 91}, 223 
 (1996).
\bibitem{Flach3} S.\ Flach, 
Phys.\ Rev.\ E {\bf 50}, 3134 
 (1994).
\bibitem{MaAub} J.\ L.\ Mar\'{\i}n and S.\ Aubry, 
Nonlinearity {\bf 9}, 1501 
 (1996).
\bibitem{DeyElFlaTsi} B.\ Dey, M.\ Eleftheriou, S.\ Flach, G.\ P.\ Tsironis, 
Phys.\ Rev.\ E {\bf 65}, 017601 
 (2001). 
\bibitem{Yuan} X.\ Yuan, 
Comm.\ Math.\ Phys.\ {\bf 226}, 61 
 (2002).
\end{thebibliography}
\end{document}